\documentclass{amsart}
\usepackage[utf8]{inputenc}
\usepackage[english]{babel}
\usepackage{amsmath}
\usepackage{setspace}
\usepackage{graphicx}
\usepackage{float}
\restylefloat{table}
\usepackage{csquotes}
\usepackage{array,multirow}

\usepackage{graphicx}
\usepackage{xcolor}
\usepackage{tikz}
\usepackage{caption}
\usepackage{subcaption}
\usetikzlibrary{arrows}
\usetikzlibrary{
    decorations.markings,
    decorations.pathmorphing,
    decorations.text,
}

\definecolor{sColour}{HTML}{F9E79F}
\definecolor{iColour}{HTML}{97BCEA}
\definecolor{rColour}{HTML}{66CDAA}
\definecolor{dColour}{HTML}{525052}
\definecolor{grayColour}{HTML}{F4F3F4}
\DeclareMathOperator*{\argmin}{arg\,min}

\title[On automatic calibration of the SIRD model]{On automatic calibration of the SIRD epidemiological model for COVID-19 data in Poland}

\author[P. Błaszczyk at al.]{Piotr Błaszczyk}
\address[P. Błaszczyk]{Faculty of Computer Science, Electronics and Telecommunications, AGH University of Science and Technology, Krakow, Poland}
\email{blaszczy@agh.edu.pl}

\author[]{Konrad Klimczak}
\address[K. Klimczak]{Faculty of Metals Engineering and Industrial Computer Science, AGH University of Science and Technology,  Krakow, Poland}
\email{klimczak@agh.edu.pl}

\author[]{Adam Mahdi}
\address[A. Mahdi]{Faculty of Applied Mathematics, AGH University of Science and Technology, Kraków, Poland; 
Oxford Internet Institute, University of Oxford, Oxford, UK}
\email{adam.mahdi@oii.ox.ac.uk}

\author[]{Piotr Oprocha}
\address[P. Oprocha]{Faculty of Applied Mathematics, AGH University of Science and Technology, Kraków, Poland}
\email{oprocha@agh.edu.pl}

\author[]{Paweł Potorski}
\address[P. Potorski]{Faculty of Applied Mathematics, AGH University of Science and Technology, Kraków, Poland}
\email{potorski@agh.edu.pl, corresponding author}

\author[]{Paweł Przybyłowicz}
\address[P. Przybyłowicz]{Faculty of Applied Mathematics, AGH University of Science and Technology, Kraków, Poland}
\email{pprzybyl@agh.edu.pl}

\author[]{Michał Sobieraj}
\address[M. Sobieraj]{Faculty of Applied Mathematics, AGH University of Science and Technology, Kraków, Poland}
\email{sobieraj@agh.edu.pl}

\markleft{P. Błaszczyk et al.}

\begin{document}

\begin{abstract}
We  propose a novel methodology for estimating the epidemiological  parameters  of a modified SIRD model (acronym of  \textbf{S}usceptible,  \textbf{I}nfected, \textbf{R}ecovered  and  \textbf{D}eceased  individuals) and perform a short-term forecast of SARS-CoV-2 virus spread. We mainly focus on forecasting number of deceased. The procedure was tested on reported data for Poland.
For some short-time intervals we performed numerical test investigating stability of parameter estimates in the proposed approach. Numerical experiments confirm the effectiveness of short-term forecasts (up to 2 weeks) and stability of the method. To improve their performance (i.e. computation time) GPU architecture was used in computations.
\end{abstract}
\maketitle

\tableofcontents

\section{Introduction}
Recently, several mathematical models have been developed to study spread of SARS-CoV-2 coronavirus and forecasting evolution of the COVID-19 pandemic, see, for example, \cite{EKuhl}, \cite{MSC_1}, \cite{VSiess}, \cite{PLOS_comp}, \cite{DIEK_1}, \cite{COMO1}, \cite{COMO2}.
Many different approaches to control and decrease number of infections were made by governments resulting, among others, in limiting the virus transmissions. 
The epidemiological situation were constantly changing due to appearance of new strains of the virus too.
Moreover, several vaccines, effective against severe disease and hospitalisation caused by the virus, were developed and delivered to society strengthening individuals immune response.
Knowledge about the transmission of infection and its evolution is an important tool, providing decision-making support to policymakers, usually governments, whose decisions are crucial in maintaining health care system stable and fully functional.

Lots of attention was also put to analyze efficiency of modelling by well established epidemiological model (SIR, SEIR, SEIRV, etc.), and especially, identification of model parameters \cite{PolCov,MumSeir,EKuhl,MSC_1,COMO1} which are crucial for proper model fitting into real data. Moreover, due to rapid increase of computational power in recent years, modern estimation methods allow exploration of model parameters as they appear and progressive update of their values. For example, in \cite{PLOS_chile} the authors compare different time series methodologies to predict the number of confirmed cases of and deaths due to COVID-19 in Chile. Bayesian approach for the agent-based model were used in \cite{PLOS_Bayesian}. Finally, deep learning methods (based on LSTM neural networks) have been investigated in \cite{PLOS_LSTM}.

In this paper we show how to estimate parameters of compartment SIRD model. This model is a modification of well known SIR model and comparing them SIRD model includes additional compartment dedicated to deceased individuals. We combine numerical method for ODEs, PSO optimization technique (particle swarm optimization, see \cite{PSO_REF}) together with the machine learning approach. The procedure was tested on the data reported for Poland. However, it can be used for any other country. Moreover, comparing to other studies, we extensively make use of GPU architecture. This makes our computations efficient and relatively fast, when comparing to analogous computations performed on pure CPU. Therefore, we can compute in a quick way many possible epidemic scenarios and compare them with the real observations.

It seems that in many cases the published data on the coronavirus pandemic are of questionable quality and its only reliable component that remains is the number of deceased individuals. Hence, as in \cite{VSiess} as the base model we chose the SIRD model, which is the extended version of the SIR model. Inclusion of the mortality data allows us to calibrate the model. Furthermore, we are aware that it seems to be impossible to calibrate this base model for the whole time interval, starting from 18 March 2020 to 10 June 2021. However, we observed that, despite of the simplicity of the SIRD model, it gives reasonable calibration/prediction results for the shorter period of time. Therefore, we propose overlapping window-wise calibration that uses a moving time windows of fixed length and estimates the unknown SIRD model parameters in each time window. In a calibration procedure we minimize suitably chosen cost functions and, due to the fact that we cannot compute gradient of the cost functions, we use PSO procedure just mentioned above.

We summarize main contributions of this paper as follows:
\begin{itemize}
    \item We propose an efficient way to automatic calibration and estimation of parameters of the chosen model.
    \item We discuss the performance of the above models with regards to high performance computing techniques (gain of GPU usage).
    \item Having defined efficient calibration and estimation procedures we compute exemplary epidemic scenarios (predictions for three-week periods) and compare them with the observed data.
\end{itemize}

The paper is organized as follows. In Section 2 data and introduction to the SIRD model with its modification is provided. Section 3 contains detailed description of the algorithm for efficient parameter estimation that has been used. Eventually, obtained results are described in Section 4.
\section{Data and model}

\begin{figure}[t]
\centering
\includegraphics[width=\textwidth]{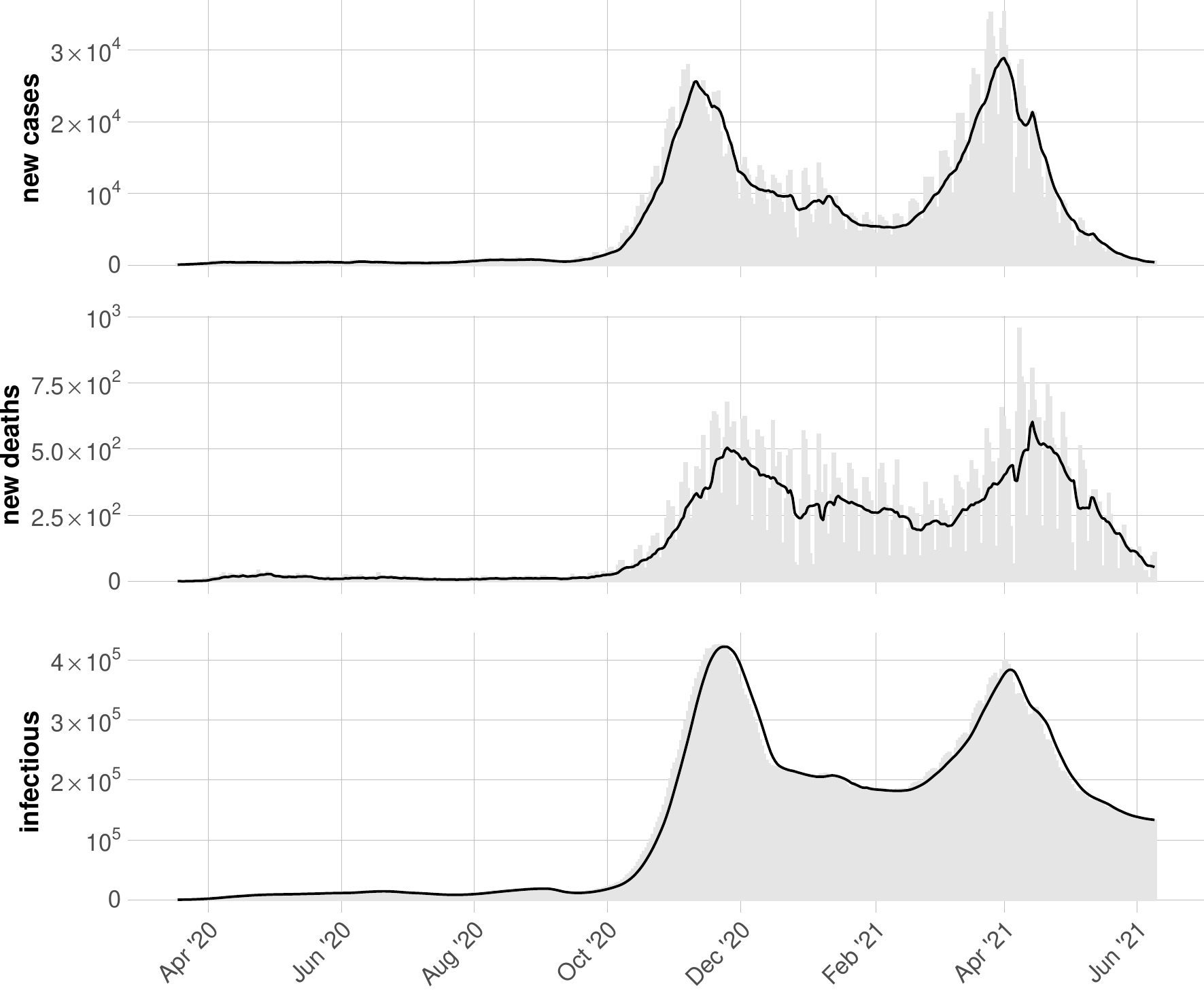}
\caption{The reported daily number of new cases, deaths and infectious individuals (gray bars) from March 2020 to June 2021. Its seven-day moving averages are depicted using black lines.}
\label{fig:reportsPoland}
\end{figure}

\subsection{Data source and preprocessing}
We extracted data from the OxCOVID19 Database \cite{Mah20}, which contains geographically unified information on epidemiology, government response, demographics, mobility and weather at a national and sub-national level collected from various sources. 


For the analysis, knowledge of daily numbers related to the virus spread is required. We created a data set containing time series describing the current number of infectious, individuals (see Figure \ref{fig:reportsPoland}), cumulative number of recovered and cumulative number of deaths due to COVID-19. 
{Specifically,  prior to analysis of the epidemiological data, missing daily values  were filled in by linear interpolation.} The new cases and deaths per day were then computed as the difference between values for successive days. {Negative values were replaced with the last non-negative observation}. Such values commonly arise when reporting authorities correct their figures for total cases or total deaths. After this initial cleaning substantial stochasticity is still present in the time series, due to factors such as backlogs in the number of cases over weekends and errors in consolidating municipal sources. To better understand the underlying trend, we computed a seven-day moving average to smooth the data (black curve on the Figure \ref{fig:reportsPoland}) as it is a common practice to account for the weekly periodicity in reporting. 

We note that the infectious individuals are those who were classified as Covid-19 positive and are currently assumed to be infectious. For the purpose of this work the infectious individuals $I(t)$ at the day $t$ are computed as follows
\begin{align*}
I(t) = I(t-1) + N_d(t) - R_d({t}) - D_d({t})    
\end{align*}
where $N_d(t)$ are the new cases detected, $R_d(t)$ are the recovered, $D_d(t)$ are the deceased at the given day $t$. 


\subsection{SIRD model}
We considered two basic compartmental models of disease transmission that can be fitted to data merging from local and national epidemiological data \cite{Tol20}.
The motivating question was to which extent can such simple models help in forecast of future evolution of daily epidemiological data.
%
%
The SIR model \cite{Ker91i,Ker91ii}, one of the simplest mathematical approaches to modelling the dynamics of infectious diseases, assumes a population of size $N$ divided 
into: $S$ susceptible, $I$ infectious and $R$ removed (immune or deceased) individuals. The three variables are time-dependent and represent the number of people in each category at a particular time. The model assumes that the deaths are subset of resistant individuals which can be estimated from the evolution of $R$ and disease does not introduce new susceptible people after recovery. The SIR model, considered on the time interval $[a,b]\subset [0,T]$ and without vital dynamics (birth and deaths), is described by the following system of equations
\[
S' = -\frac{\beta}{N}SI, \qquad
I' = \frac{\beta}{N}SI-\gamma I,\qquad
R' = \gamma I\\
\]
where $t\in [a,b]$ and $\beta$ is the transmission rate (controls the rate of spread representing the probability of transmitting disease between a susceptible and an infectious individual), $\gamma$ is the recovery rate constant and $R_0=\beta/\gamma$ is the basic reproduction number. 
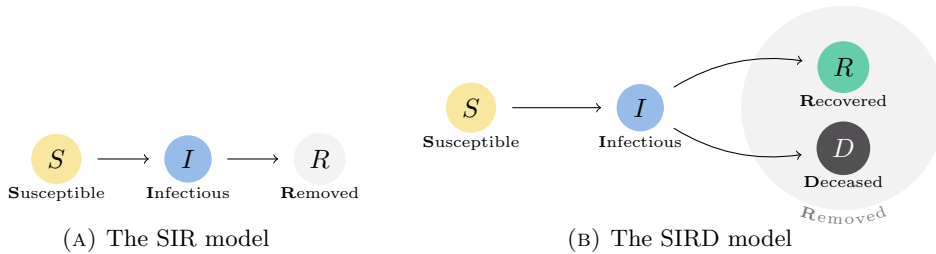
\begin{figure}
\begin{subfigure}[b]{0.35\textwidth}
\centering
\begin{tikzpicture}[scale=0.7]
\node[shape=circle,thick,fill=sColour] (S) at (0.5,0) {\bf{$S$}};
\draw (S) node[below,yshift=-7pt] {\tiny{{\bf{S}}usceptible}};
\node[shape=circle,fill=iColour] (I) at (3,0) {\bf{$I$}};
\draw (I) node[below,yshift=-7pt] {\tiny{{\bf{I}}nfectious}};
\node[shape=circle,fill=grayColour] (R) at (5.5,0) {{\bf{$R$}}};
\draw (R) node[below,yshift=-7pt] {\tiny{{\bf{R}}emoved}};
\path [->,shorten <=2mm,shorten >=2mm] (S) edge (I);
\path [->,shorten <=2mm,shorten >=2mm] (I) edge (R);
\end{tikzpicture}
\caption{The SIR model}
\label{fig:sir}
\end{subfigure}%
\begin{subfigure}[b]{0.73\textwidth}
\centering
\begin{tikzpicture}[scale=0.9]
\filldraw[gray!10] (6,0) circle (1.5);
\node[shape=circle,thick,fill=sColour] (S) at (0.5,0) {\bf{$S$}};
\draw (S) node[below,yshift=-7pt] {\tiny{{\bf{S}}usceptible}};
\node[shape=circle,fill=iColour] (I) at (3,0) {\bf{$I$}};
\draw (I) node[below,yshift=-7pt] {\tiny{{\bf{I}}nfectious}};
\node[shape=circle,fill=rColour] (R) at (6,0.6) {\bf{$R$}};
\draw (R) node[below,yshift=-7pt] {\tiny{{\bf{R}}ecovered}};
\node[shape=circle,fill=dColour] (D) at (6,-0.6) {\textcolor{white}{\bf{$D$}}};
\draw (D) node[below,yshift=-7pt] {\tiny{{\bf{D}}eceased}};
\draw [thin,grayColour,postaction={decorate, decoration={text align={center},raise={-2mm},text along path, text={|\tiny\color{gray}|{\bf{R}}emoved}}}] (4.5,0) arc (180:360:1.5cm);
\path [->,shorten <=2mm,shorten >=2mm] (S) edge (I);
\path [->,shorten <=2mm,shorten >=2mm] (I) edge[bend left=20] (R);
\path [->,shorten <=2mm,shorten >=2mm] (I) edge[bend right=20] (D);
\end{tikzpicture}
\caption{The SIRD model}
\label{fig:sird}
\end{subfigure}%
\caption{A schematic representation of the SIR and the SIRD model.}
\label{fig:models}
\end{figure}
%
%
%
%
%
Many countries report not only daily number of new positive cases and change of infectious individuals but also number of deaths due to COVID-19. It could be seen as a valuable data. The reported data describing number of new positive cases are very often underestimated, what could be caused by the testing approach in the given country, whereas it is less likely to happen in case of number of deaths.  Namely, severe patients who were tested and hospitalised before they die are always included in reports. Taking deaths into consideration as a separate compartment in a model give the possibility to find out more about the dynamics of the pandemic.
This is the main motivation to extend the SIR model slightly, under hypothesis that it will make our modelling more reliable and describing more aspect of the disease. 

The SIRD model aims to differentiate between the Recovered (i.e. individuals who survived and are now immune) and the Deceased. The model equations without vital dynamics are
\begin{equation}\label{eq:SIRD}
S' = -\frac{\beta}{N}SI,\qquad
I' = \frac{\beta}{N}IS - (\gamma+\mu) I, \qquad
R' = \gamma I,\qquad
D' = \mu I,
\end{equation}
where $\beta$, $\gamma$, and $\mu$ are the rates of infection, recovery, and mortality, respectively. From this moment everywhere we mention $S,I,R,D$ compartment or a value of one of the following parameters $\beta$, $\gamma$, and $\mu$ we refer to the compartments and the parameters of the SIRD model.
%
%
Considering the SIRD model on the time interval $[a,b]\subset [0,T]$
we allow the transmission rate $\beta$ in the SIRD model to be time varying
\begin{align}\label{eq:beta}
\beta(t)=\begin{cases}
\beta_1,  &t\in [a,t_1), \\
\beta_1+\frac{\beta_2-\beta_1}{t_2-t_1}(t-t_1), &t\in [t_1,t_2),\\
\beta_2,  &t\in [t_2,b],
\end{cases}
\end{align}
with $t_1,t_2\in [a,b]$. The reason for this approach is that there are several governmental policies (social distancing, face masks, closed schools, etc.), the main objective of which is to change the speed of spreading a virus. Moreover, the appearance of new and more transmissible  strains of the virus caused different dynamics of the pandemic. This stays in contradiction to the assumption of  non-time-varying parameters of the standard SIRD model and needs to be included  in a suitable way. On the other hand, we believe that in short time intervals, coefficient $\beta$ cannot change too much (cf. \cite{VSiess}); 
see also pages 128-129 in \cite{EKuhl} where a hyperbolic tangent type ansatz for the transmission rate $\beta$ was considered. Time-dependent $\beta$'s of such types allow us to model effects of government interventions.

For the SIRD model the basic reproduction number
could be calculated using formula  $R_0=\frac{\beta}{\gamma+\mu}$.  This number could be seen as indicator of pandemic phase. If it is greater than one it means that the virus is spreading in a population, if it is less than one the outbreak is fading out.


\section{Efficient parameter estimation}

In this section we will present how parameters in \eqref{eq:SIRD} together with \eqref{eq:beta} can be effectively estimated. As we mentioned, formulas \eqref{eq:beta}
were introduced under assumption that our observations cover sufficiently short period of time. Therefore, instead of modelling of the whole period when data was measured,
we will focus on smaller time windows when the model will be executed. This way we will obtain several different approximations with possibly different parameters.
One of the aims of this study will be analysis of differences between these parameters, their stability and most importantly, utility of this approach in
short terms forecasting of possible future evolution of parameter $D$ derived from the model.
\subsection{Model calibration}
As we explained earlier, in our approach we will divide considered time interval of data $[0,T]$ into smaller time windows, where simulations will be executed.
We define these windows by
 $I_i=[T_i,T_i+\tau]$
such that
\begin{align*}
T_{i+1}=T_{i}+\delta
\end{align*}
where the first window starting point is $T_1=0$, the length of the window $\tau$ is fixed and $\delta$ is a shift of the window. Since we have daily data $T$, $\delta$ and $\tau$ are always integers. Moreover, in this case we constructed sequence of windows, consisting of next $\tau+1$ observations that have been taken from the initial sequence of observations, by moving the starting point of the next window by $\delta$ days comparing with the latter. This way we obtain $\left \lfloor{1+(T-\tau)/\delta}\right \rfloor$ windows where fitting takes place.

For our tests $T$ was fixed to $449$, that is we cover $450$ days of data, starting from Mar 18, 2020 to Jun 10, 2021. We set $\delta$ to $3$ days and $\tau=35$ (10\% of yearly data),
resulting in $139$ windows. As we can see, numerous windows are overlapping, and for close indices $i$ these overlaps are huge. This intuitively suggests, provided that the model parameters are stable,
that close windows will have close parameters and some kind of continuous change of parameters will occur during fitting process.


Next, let us explain our approach to estimation of the parameters of the model.
When fitting the SIRD model, in order to estimate the unknown parameters, we may consider all three variables $R,I,D$ since simultaneously all of them were present in the dataset. 
Based on these variables we distinguished two different approaches.

The first one is based on the assumption that our main objective function depends on a single compartment from the model.  
Assuming that the sequence $\{y_{i}\}_{i=1}^{n}$ contains reported values of the single compartment $Y \in \{I, R, D\}$, for each consecutive day from $1$ to $n$, and $\{\widehat{y}_i\}_{i=1}^{n}$ refers to the corresponding values obtained from the Euler scheme (applied to \eqref{eq:SIRD}) with initial condition $\widehat{y}_1 = y_{1},$  we introduce the following low-level cost functions:
\begin{align*}
\text{MXSE}(Y) &= \max_{i=1,\dots,n}e_i^2 && \text{max squared error} \\
\text{MSE}(Y) &= \frac{1}{n}\sum_{i=1}^{n}e_i^2 && \text{mean squared error}  \\
\text{MAE}(Y) &= \frac{1}{n}\sum_{i=1}^{n}|e_i| && \text{mean absolute error}\\
\text{MAPE}(Y) &= \frac{100\%}{n}\sum_{i=1}^{n}\left |\frac{e_i}{y_i}\right| && \text{mean absolute percentage error}
\end{align*}
where
$$e_i=y_i-\widehat{y}_i\qquad  \mbox{for} \ i=1,\dots,n.$$ 
Hence, the main cost function for a single arbitrarily chosen compartment is defined as one of the previously introduced low-level functions with respect to the specified compartment. 
For the first part of further research we chose compartment $D$, since it seemed to exhibit a significant consistency with the real-world data reports. As a result we ponder four cost-functions and defined them using the following notation  
\begin{align*}
F^C_D=C(D)\qquad \mbox{for }C \in \{\text{MXSE, MSE, MAE, MAPE}\}.
\end{align*}
For chosen function type $C$ the main objective is to find parameters that minimize the value of $F^C_D$, that is
\begin{align*}
\argmin_{\beta_1,\beta_2, t_1, t_2, \gamma, \mu} F^C_D.
\end{align*}

The second approach is based on the assumption that the main objective function depends on all three compartments, namely $I, R, D$. It is worth noticing that the second approach is by far more demanding when it comes to computations.

Since every single compartment of the model might have values from different ranges, compared to each other, the error values of introduced low-level cost functions, excluding MAPE case, may strongly differ. Therefore, to treat every compartment equally, we simply use a~proper normalization. Let us introduce the following family of functions
$$
f_{Y}(y) = \frac{y - \min\limits_{i=1,...,n}y_{i}}{\max\limits_{i=1,...,n}y_{i} - \min\limits_{i=1,...,n}y_{i}}
\qquad \mbox{where} \ Y \in \{I, R, D\}
$$
and $\{y_{i}\}_{i=1}^{n}$ denotes a sequence of reported values of $Y$ from the consecutive days $\{1,...,n\}$. The introduced functions were used to rescale not only reported values but also values obtained from the Euler scheme for compartments $I, R, D$. This was in order to make possible to compare among themselves. In particular, this is how we created, based on the sequence $\{\widehat{y}_{i})\}_{i=1}^{n}$, a new one $\{f_Y(\widehat{y}_{i})\}_{i=1}^{n}$ that we associate with the variable $\widetilde{Y}$.
Consequently, to take advantage of rescaling, that is required for every low-level cost function beside $\text{MAPE}$, the values of $e_{i}$ were replaced by $\tilde{e}_{i} = f_{Y}(y_{i}) - f_{Y}(\widehat{y}_{i})$, that is replaced by errors for compartment $\widetilde{Y}$. 
The family of main objective functions consists of functions that are referred to as the maximum of low-level cost functions, i.e
\begin{align*}
F_{IRD}^C = \max\{C(\widetilde{I}), C(\widetilde{R}), C(\widetilde{D})\}\qquad \mbox{where } C \in \{\text{MXSE, MSE, MAE}\}
\end{align*}
and
\begin{align*}
F_{IRD}^{\text{MAPE}} = \max\{C({I}), C({R}), C({D})\}
\end{align*}
as it do not need normalization.
In such a way we obtain four more different cost functions, namely
\begin{align*}
F_{IRD}^{\text{MXSE}}, \ F_{IRD}^{\text{MSE}}, \ F_{IRD}^{\text{MAE}}, \ \mbox{and} \ F_{IRD}^{\text{MAPE}}.
\end{align*}
Each of these values might be used independently as an indicator for finding the best fit to the model. 
Similarly to the previous group of cost functions in this case for chosen function type $C$ the main aim is to find parameters that minimize the value of $F^C_{\text{IRD}}$
\begin{align*}
\argmin_{\beta_1, \beta_2, t_1, t_2, \gamma, \mu} F_{IRD}^{C}.
\end{align*}

To evaluate accuracy of the fitting and compare performance of considered objective functions we need a benchmark function, whose formula is in some sense independent
of objective functions. Otherwise, we would give preference to one of them. In our experiments we decided to consider an $R^{2}$ coefficient for measuring the fitting accuracy independently of defined low-level cost functions
\begin{align*}
R^2(Y) &= 1 -  \frac{\sum_{i=1}^{n} e_i^2}{\sum_{i=1}^{n} (y_i - \bar{y})^2} && \text{coefficient of determination}
\end{align*}
which was always calculated based on variable $Y=D$ only.
While we use different objective functions, our ultimate goal is the best possible forecast of the compartment $D$. That is the main reason to focus on this compartment of the model.

\subsection{Preprocessing}

Before running the fitting procedure we have chosen the ranges of parameters that will be tested. We did it in two steps. Firstly, we choose the initial ranges of parameters. The decision was made in accordance with the epidemiological meaning of the parameters and rough estimates of them presented in many papers related to COVID-19. Nevertheless, it was supposed to be rather a~wide set of choice. Then, as we measure the goodness of fit with $R^2(D)$ we check its values for the obtained cases and observed that it is reasonable to narrow intervals of tested parameters even more.  As a~result, we obtained much better values of $R^2(D)$, as we expect. In this subsection we deliver the details of this preprocessing and received results. 

For the first model calibration procedure in a given window $I_i$ we choose to search the appropriate parameters $\beta_1,\beta_2,\gamma,\mu$ in $[0,10]$ and $t_1, t_2$ satisfying $T_{i} \leq t_1\leq t_2 \leq T_{i} + 35$. In the two first rows of Table \ref{tab:comparison} one can find mean values of $R^2(D)$ - measure applied to the compartment $D$ for each of the above objective functions, averaged over all $138$ windows for all considered objective functions. This mean could be seen as a indicator determining the goodness of fit.

\begin{figure}
\centering
\includegraphics[width=12cm]{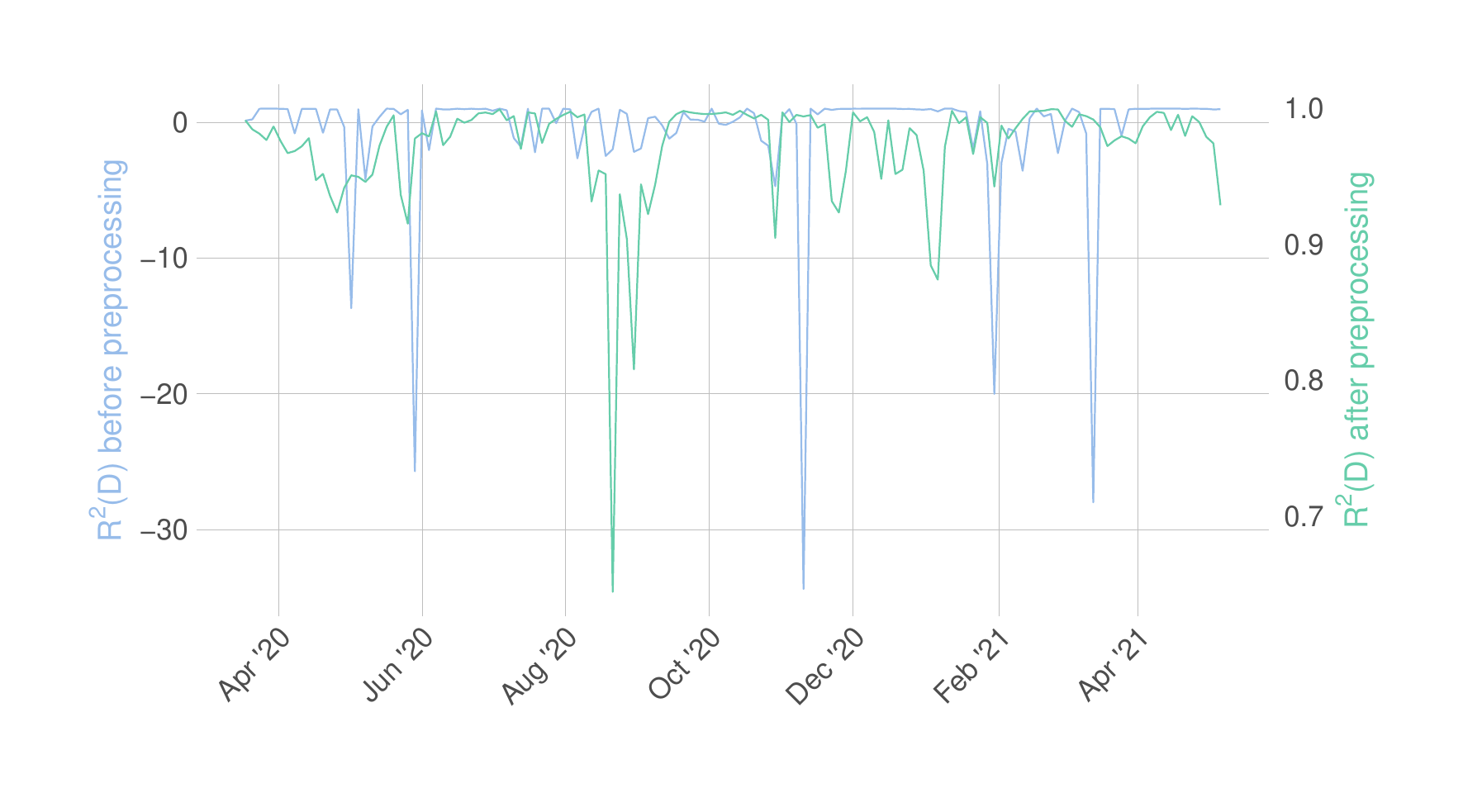}
\caption{Comparison of values $R_0^2(D)$ obtained by cost-function $F_{IRD}^{\text{MXSE}}$ before (left values axis) and after preprocessing (right values axis).}
\label{fig:comparison}
\end{figure}

Based on the obtained parameters values from initial run and its epidemiological meaning in the model we decided to restrict the parameters bounds even more and rerun the procedure in order to obtain more accurate results. This time we choose $\beta_1,\beta_2 \in [0, 2], \ \gamma \in [0, 1], \ \mu \in [0, 0.1]$. Moreover, we made one more change regarding $t_1,\ t_2$ namely, we assume that $T_i \leq t_1 \leq t_2 \leq T_i+35-7$. It is due to our main forecasting approach that is based on the extrapolation of the fitted model. Therefore, the obtained values of the parameters $\beta_{2},\ \gamma$ and $\mu$ are crucial. On the other hand, the fitting procedure may lead to the case when $t_{2}$ is very close to $\tau$. Hence, $\beta_{2}$ may become irrelevant by having arbitrary value regardless of the quality of the fit, which may result in unreliable forecast. To prevent from this we decrease the upper bound for $t_{1}$ and $t_{2}$ by $7$ days. As before, this time we evaluate mean values of $R^2(D)$ too. One could find them in last two rows of Table \ref{tab:comparison}. Moreover, for $F_{IRD}^{\text{MXSE}}$ more detailed outcomes of $R^2(D)$ that is its values for all considered windows before and after preprocessing are presented in Figure \ref{fig:comparison}. On this graph every value of $R^2(D)$ is associated with the starting day of a window for which it was calculated. As it can be seen in the table and figure, after narrowing the search space of the parameters, the fitting procedure resulted in enhanced performance. Similar results are observed for all considered functions. 

For all given approaches the mean values of $R^2(D)$ after preprocessing are very close to each other and almost indistinguishable. The only value that is significantly different from the others is that for MAPE when we fit using all three variables.This is probably because MAPE severly punishes errors made for small values while neglecting errors made for great values. This is different from other metrics used here as well as the scoring function, which hold absolute values in great importance. Considering the order of magnitude of fitted variables it is reasonable for MAPE to yield worse results. As a representative cost-function for further investigations (and figures we produce) we chose $F_{IRD}^{\text{MXSE}}$ because it gave the best fit to $D$ from all functions $F_{D}^C$ and $F_{IRD}^C$.

\begin{table}[H]
\begin{tabular}{llrrrr}
\hline 
procedure run &\text{fitting to} & MXSE & MSE  & MAE & MAPE \\
\hline \hline
\multirow{2}{*}{before preprocessing} &$D$ only & 0.86361  & 0.98422   & 0.98594 & 0.98420\\ \cline{2-6}
&$I, R, D$ & $-0.65582$  & $-9.85391$  & $-8.36460$ & $-20.42675$ \\
\hline \hline
\multirow{2}{*}{after preprocessing} &$D$ only & 0.99932  & 0.99951  & 0.99956 & 0.99937\\ \cline{2-6}
 &$I, R, D$ & 0.97289  & 0.95640  & 0.97282 & 0.89414\\
\hline
\end{tabular}
\caption{Objective functions comparison - mean values of $R^2(D)$ over all considered windows calculated based on variable $D$ only.}
\label{tab:comparison}
\end{table}



\subsection{Efficient computational issues}

Since the exact solution of SIRD model is unknown, the gradient methods used for optimization are not applicable. Therefore, we use well known Particle Swarm Optimization algorithm  (abbrev. PSO), see, for example, \cite{PSO1}. The method does not require knowledge of the gradient values.
The base coefficients were set as follows:

\begin{enumerate}
    \item Number of particles: 10000
    \item Innertia coefficient: 0.5
    \item Social coefficient: 0.5
    \item Cognitive coefficient: 0.5
    \item Maximal number of iterations: 100
\end{enumerate}

The boundary problems ensued by algorithm proposing out-of-bounds solutions were solved by setting the out-of-bounds values to boundary values. \\

Since every particle state of the PSO method can be calculated independently, taking advantage of the parallel computations is crucial when it comes to high performance.
Nowadays, this is where the modern Graphics Processing Units (GPUs) play a key role due to huge number of parallel processors (cores) compared to the classical Central Processing Units (CPUs).
Therefore, the first most demanding part of the algorithm that consist of finding the approximate solution of every single particle ODE (SIRD Ordinary differential equation with certain set of parametrs), was performed on GPU, namely Nvidia's Titan V and GeForce RTX 2080. However, the second part that consists of the error calculation was performed on CPU. In fact it gives a room for possible future improvements. 

The code was fully written in Python programming language with the usage of the most common third-party libraries like NumPy and SciPy.
In order to boost the application performance the CUDA architecture was also employed via third-party library and just-in-time compiler named Numba. From the
numerical point of view the classical Euler method was embedded in order to find the approximated solutions of the introduced system of ODEs.

In order to validate GPU performance over CPU, we managed to run 20 independent simulations consisting of single PSO iteration both on CPU and GPU. Simulations were performed for data from 4 April 2021 to 5 July 2021, which was chosen arbitrarily, and the cost function $F_{IRD}^{\text{MXSE}}$. For the performance comparison, we took average values of the obtained execution times for benchmarked devices. With the usage of Nvidia TITAN V we obtained average execution time of 2.176s with standard deviation equal to 0.175s. On the other hand, executing our code on Intel(R) Xeon(R) CPU E5-2650 v4 @ 2.20GHz and using standard and well-known SciPy function odeint from the package integrate, it took in average 55.655s with standard deviation 0.631s.

Based on the obtained results it is clear to see the reasoning behind GPU usage which can significantly decrease execution time.
\section{Results}

In this section we will present the results of fitting procedure for Poland. We can divide them into two parts. The first one is dedicated to window-wise fitting procedure obtained for reported data for Poland from 18 March 2020 to 10 June 2021. The latter is related to the results when multiple repetition of the fitting procedure for two chosen windows were performed. In the latter case we mention the extensions of the model. In both cases we constructed several graphs presenting the results such as predicted parameters and compartments values for a representative cost function, namely for $F_{IRD}^{\text{MXSE}}$.

\subsection{Window-wise calibration}
As we already mention the first part of figures summarizing the results is showing the performance of the fitting procedure. We constructed two groups of graph. First is related to the fitted parameters whereas the second to compartments values.





\begin{figure}
\includegraphics[width=\textwidth]{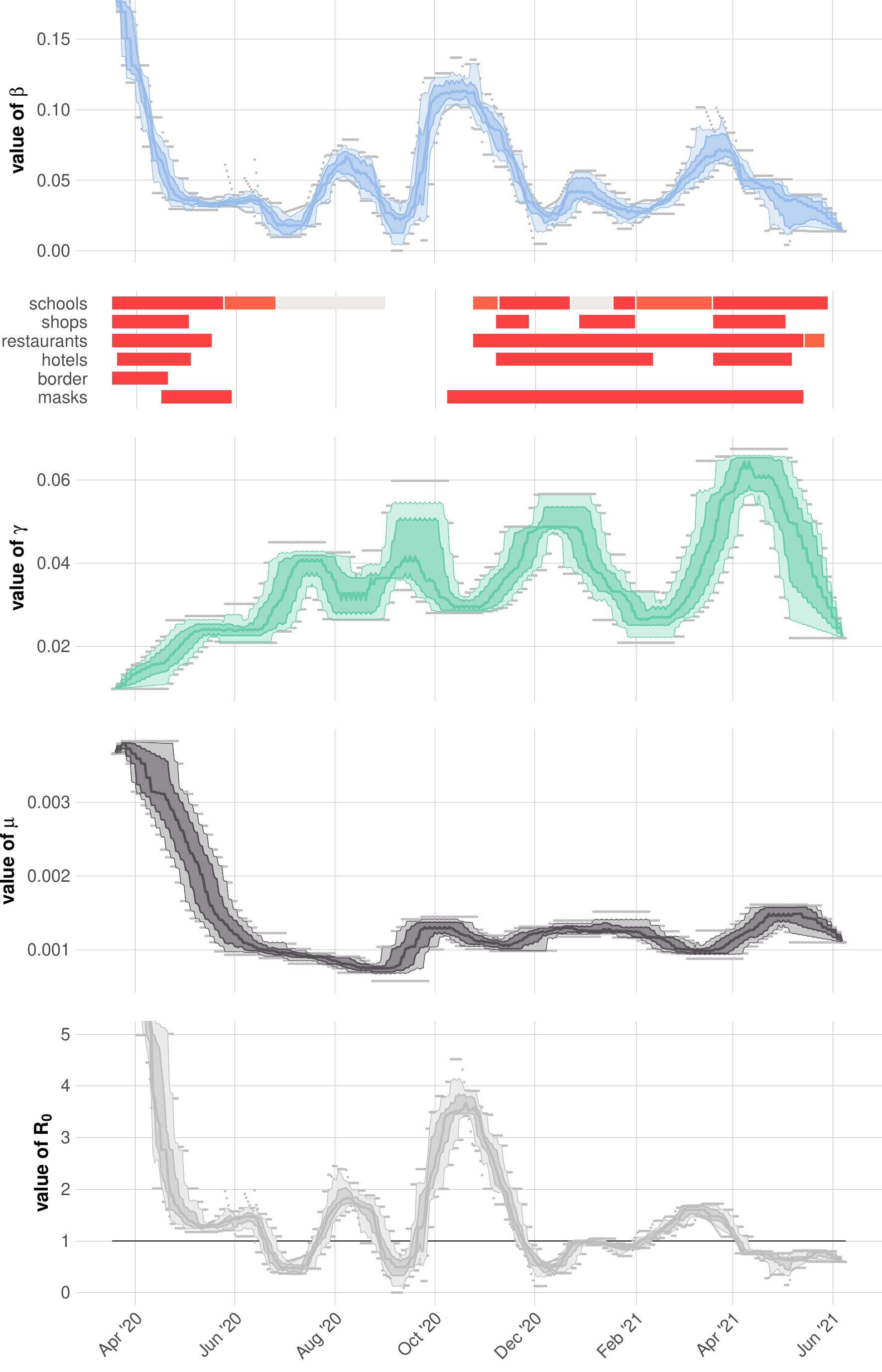}
\caption{Summary of SIRD model parameters and the based reproduction number $R_0$ based on results for cost function $F_{IRD}^{\text{MXSE}}$ with an overview of interventions introduced by the Polish Government.}
\label{fig:parametersBounds}
\end{figure}

On Figure \ref{fig:parametersBounds} we present summary of parameters and give an outline of interventions introduced by the government. To be exact, this figure presents values of the parameters obtained from different windows for a given day. It is worth mention that parameter $\beta$ is the only one that is time-dependent. Nevertheless we decided to plot all graphs in the same manner. Namely, we always consider the values of a given parameter for every day. Since windows overlap we received more than only one value for a given day. The smallest and the largest values were depicted using gray dots. Other values create the first level bound (brighter interval). Creating the next bound (inner and darker interval) we exclude two largest and two smallest values. The solid line inside represent the median for a given day. This construction allow us to observe the stability and change of parameters in time when we move slightly the starting point of the window. Using the parameters values we were able to calculate the basic reproduction number $R_0$ evaluated for each day. Its graph is added as the last one. Moreover, to give the reader a condensed overview of interventions introduced by the Polish Government in the second part of Figure \ref{fig:parametersBounds} there are indicated, using red bars, time frames when the main interventions were introduced. Periods when schools remained closed are indicated at the very beginning. For this intervention we additionally indicate the school holidays using gray bars. Next, there are presented shopping centers, restaurants, hotels and border closure as well as when face mask wearing outside was obligatory (the rule to use them indoors was unchangeable valid that time). All plots share common time axis added to the last graph.

\begin{figure}
\centering
\includegraphics[width=\textwidth]{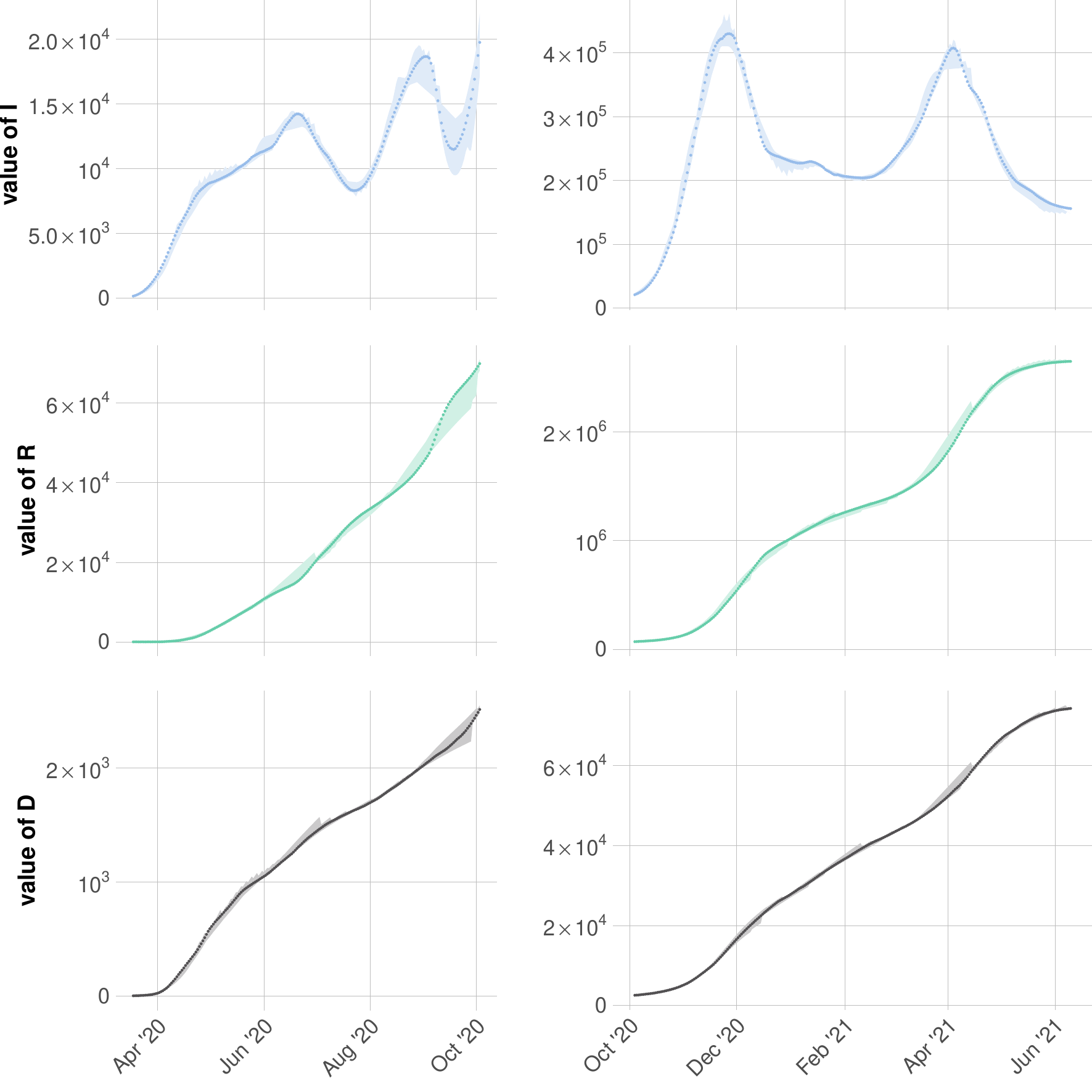}
\caption{The fitting bounds for model compartments $I$, $R$ and $D$ (brighter color envelope) obtained by $F_{IRD}^{\text{MXSE}}$ with the reported data (dots).}
\label{fig:fittingBounds}
\end{figure}

On Figure \ref{fig:fittingBounds}, where we can see the values of compartments $I,R,D$, reported data and created fitting bounds are presented. In every 36-days window using predicted parameters for the SIRD model with the initial reported data (first day of the window) we generated compartments $I$, $R$ and $D$. Because we repeat this process but considering in the next step the new window starting only three days later we get more compartments values for a given day as the windows overlap. Thus we are able to construct fitting bounds for a compartment. We mean by that the interval consisting of the minimal and maximal value of a given compartment at a given day. For a given day we choose this values from the outputs for different windows containing this day as the moving windows overlap. It is depicted as brighter area in the appropriate color related to the compartment. In the same plot the reported data are pointed using dots. Moreover, to improve legibility every graph is dividend into two parts. All plots in the same column share the common time axis added to the last graph.

\subsection{Repetitions of calibration and model extensions}

In order to check the stability of the calibration method we used we repeated the process $1000$ times for two arbitrarily chosen windows. As a result we received $1000$ sets of parameters for each window. Gathered results allow to consider how far the parameters obtained in the next repetitions of the fitting procedure are for a given window. When we add to this the observation that parameters do not change drastically in short period time we are able to
construct extensions of the model, in our simulation for the next three weeks. Consequently, based on multiple sets of parameters every of which gave a good fitting performance to the data in the window we construct $1000$ extensions creating bounds for possible values of compartments in the model using the SIRD model with parameters $\beta_2$, $\gamma$ and $\mu$ for the next weeks. Eventually, we compared how far from the reported data such short-period forecast was and discussed the efficacy of this approach.

Similarly to the previous subsection, first we present plots related to parameters and next those related to model compartments with their extensions as described.

For the analysis we chose two windows - the first from 10 May 2020 to 13 July 2020 and the second from 4 April 2021 to 08 May 2021. It is worth mentioning that every graph presenting the results is divided into two columns to give the opportunity to compare. The left one is always associated with the first window and the right one with the latter. 
These windows were taken into consideration as representatives of two different pandemic states. The first window characterise rapid change in compartment $I$ outside the observation window whereas in the second case volatility of all three variables $I, R, D$ was small. Moreover, worth mentioning is that using every function $F^C_D$ for fitting of parameters yields worse forecasting results than fitting using all three variables $I,R,D$ that is with function $F^C_{IRD}$.

\begin{figure}
\centering
\begin{subfigure}[b]{0.495\textwidth}
\includegraphics[width=\textwidth]{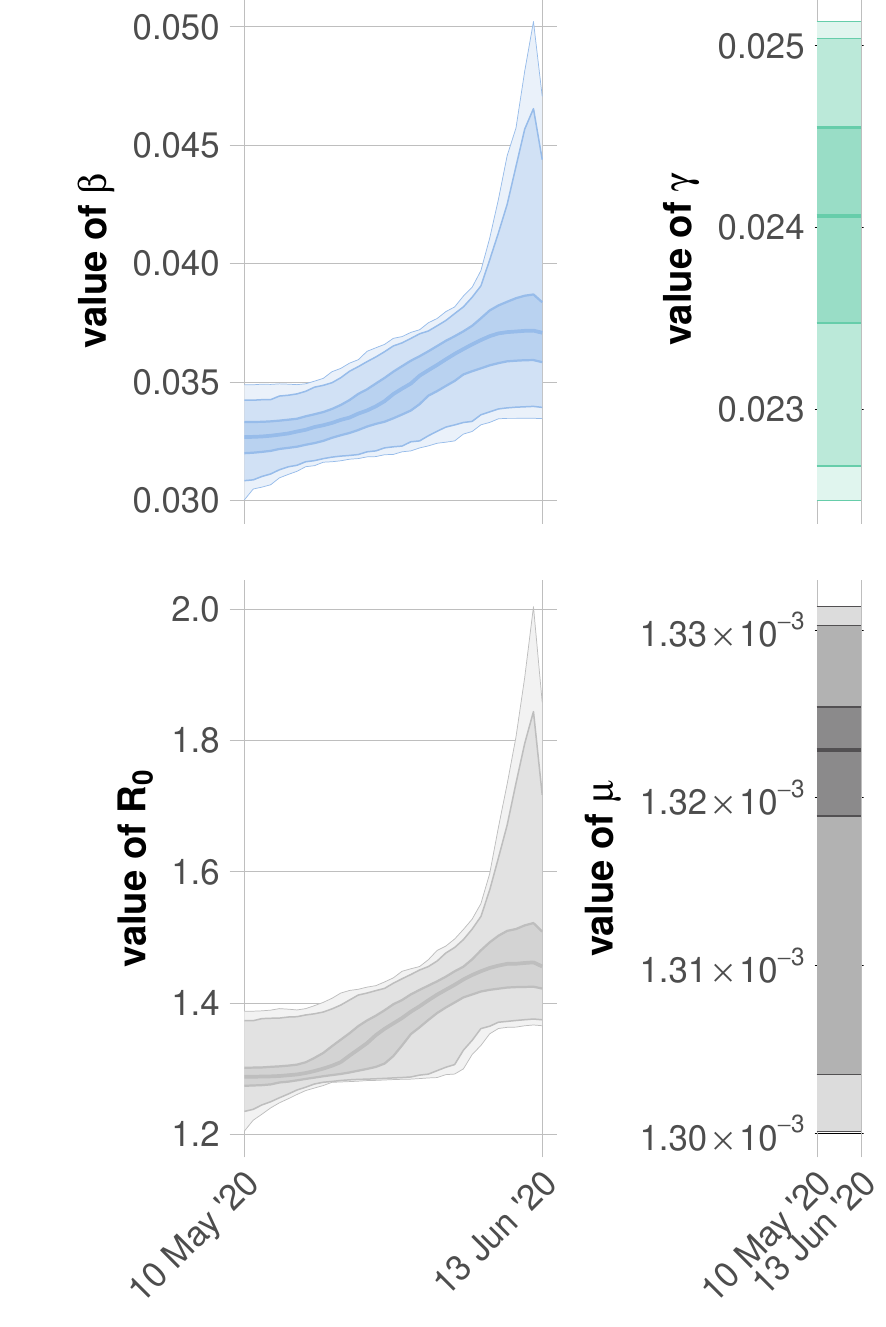}
\end{subfigure}
\begin{subfigure}[b]{0.495\textwidth}
\includegraphics[width=\textwidth]{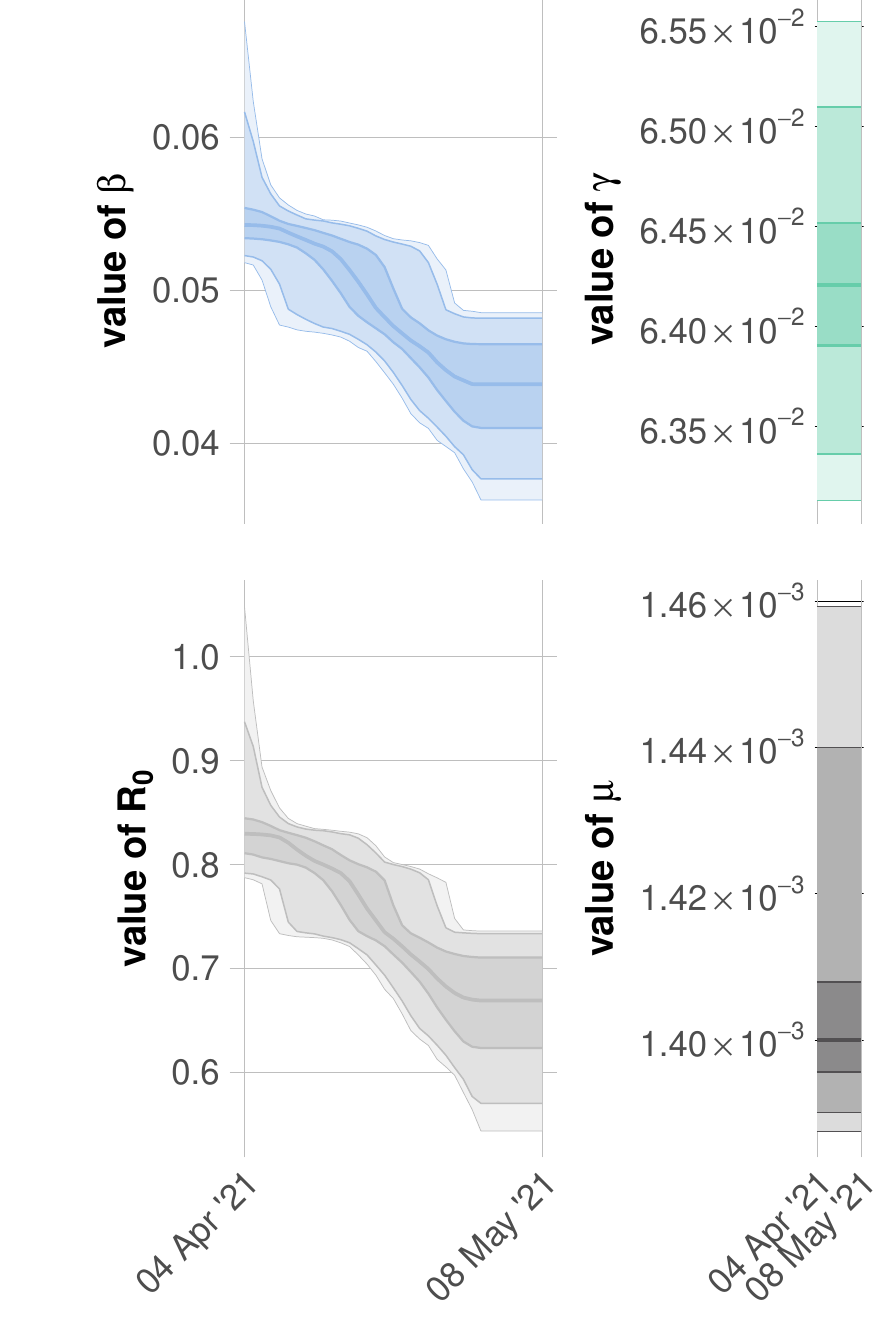}
\end{subfigure}
\caption{Parameters and their distribution based on 1000 repetition of fitting process for two windows - from 10 May 2020 to 13 July 2020 (left) and from 4 April 2021 to 08 May 2021 (right).}
\label{fig:parameter1000}
\end{figure}

Let us now come back to the question, how far the parameters obtained in the next repetitions of the fitting procedure for a given window are. On Figure \ref{fig:parameter1000} are shown some graphs presenting parameters values and their distribution. It is based on 1000 repetition of fitting process for two windows we just mentioned. 
Creating those plots we were using the idea of a boxplot construction.
For every parameter we attached plot including bounds containing $95\%$, $90\%$, $50\%$ values surrounded the median value (solid line in the middle) symmetrically. The darker the band is the less values it contains. Since $\beta(t)$ is the only parameter that is time-dependent we present its values in this manner separately for every day from considered in the window. The construction of plot for $R_0$ is similar to this for $\beta(t)$ as $R_0$ is dependent on $\beta(t)$ value. In the model there are present parameters $\gamma$ and $\mu$ but they aren't time-dependent therefore we simplify graphs presenting their distributions.

\begin{figure}
\centering
\begin{subfigure}[b]{0.495\textwidth}
\includegraphics[width=\textwidth]{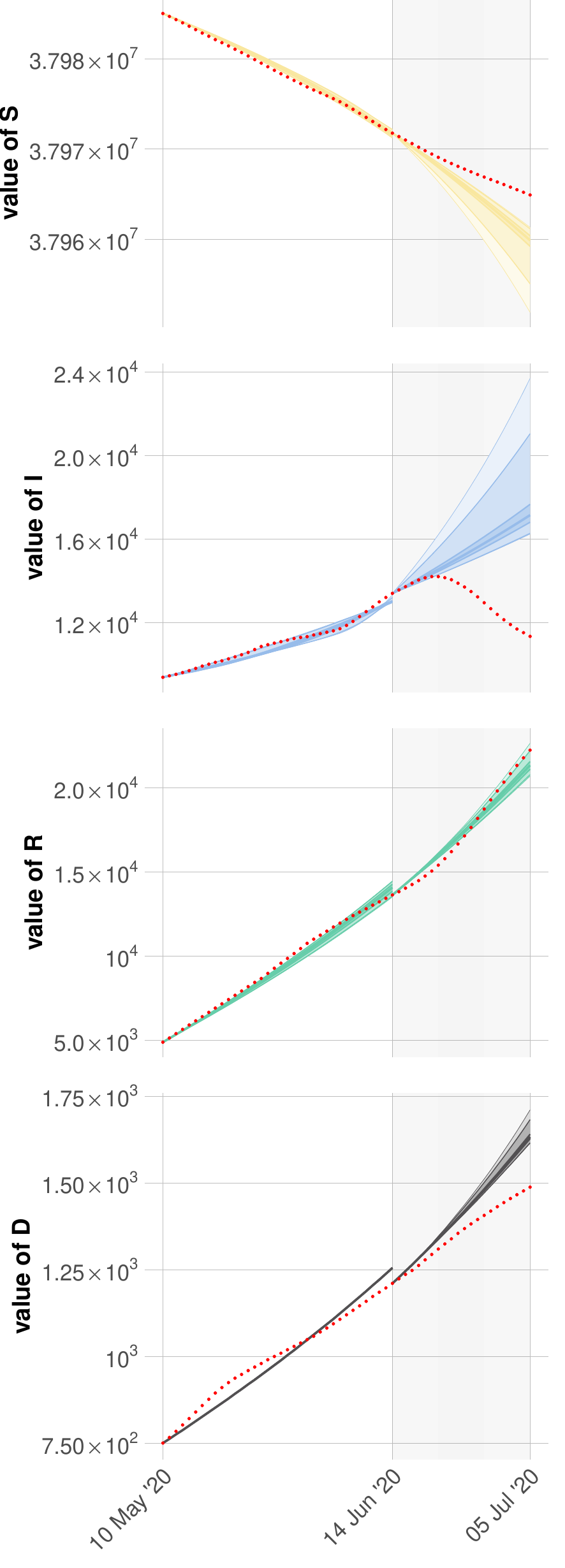}
\end{subfigure}
\begin{subfigure}[b]{0.495\textwidth}
\includegraphics[width=\textwidth]{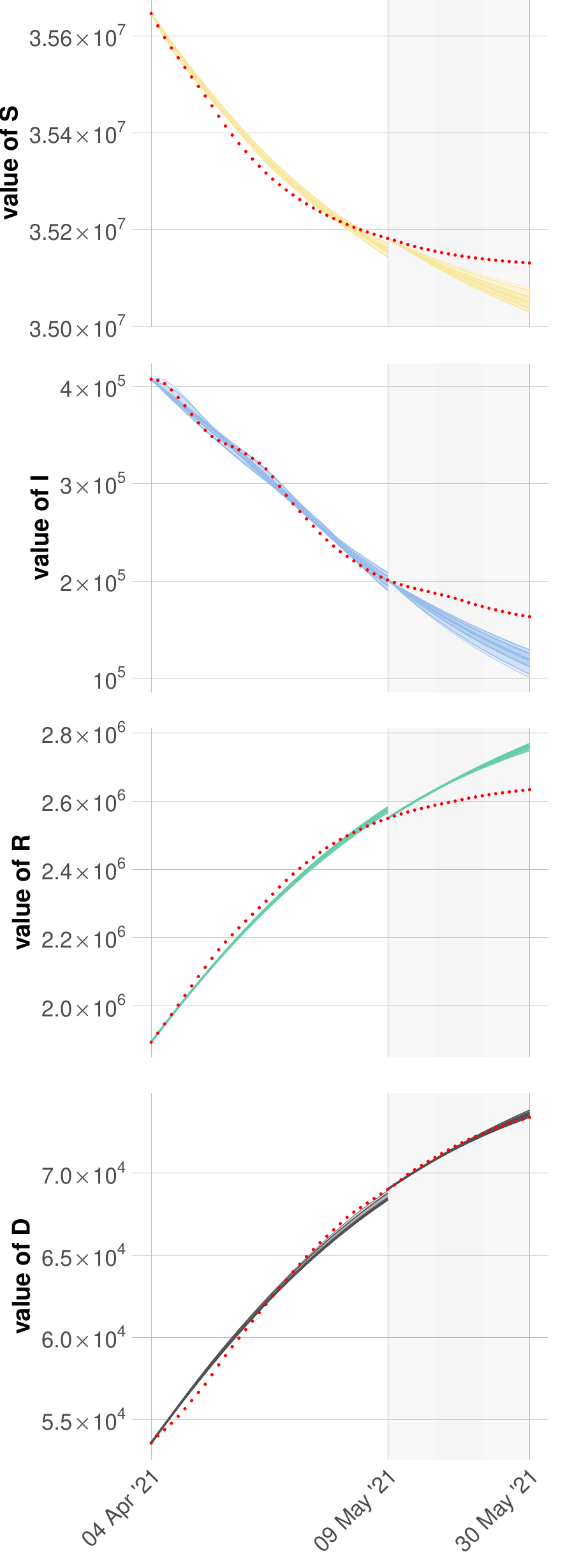}
\end{subfigure}
\caption{Compartments of the SIRD model and its extensions - values and their distribution based on 1000 repetition of fitting process for two windows - from 10 May 2020 to 13 July 2020 (left) and from 4 April 2021 to 08 May 2021 (right).}
\label{fig:compartmentsExt1000}
\end{figure}

On Figure \ref{fig:compartmentsExt1000} we collected the plots presenting the SIRD model compartments obtained by the fitting procedure with parameters we just discussed.
On this figure compartments bounds and some statistics prepared for both considered windows are presented. The manner of creating the plots is the same as for the parameters. Additionally, on the gray background one can find the model extension for the next 21 days assuming that to its construction the initial value and the parameters values were chosen from the last day of the obtained in the fitting procedure window. To give an opportunity to verify how far from the reported values not only the fitted compartments but also compartments extensions were the latter were indicated using red dots.



\section*{Acknowledgement}
The authors thank all members of the COMO Consortium and their collaborative partners for all comments and suggestions on the underlying approach used. Moreover, the investigators acknowledge the philanthropic support of the donors to the University of Oxford's COVID-19 Research Response Fund.
\\ \\
{\bf Data availability} The datasets generated and analysed during this study are available from the corresponding author on a reasonable request.
\bibliographystyle{plain}
\bibliography{cmmc}
\end{document}